\documentclass[12pt,a4paper,final]{iopart}
\expandafter\let\csname equation*\endcsname\relax
\expandafter\let\csname endequation*\endcsname\relax
\usepackage{iopams}
\usepackage{graphicx}
\usepackage{subfigure}
\usepackage[breaklinks=true,colorlinks=true,linkcolor=blue,urlcolor=blue,citecolor=blue]{hyperref}
\usepackage{cite}
\usepackage{braket}
\usepackage{mathtools}

\usepackage[a4paper]{geometry}
\usepackage{bpchem,upgreek}
	\usepackage{amsmath}
	\usepackage{makeidx}
	\usepackage{amsfonts}
	\usepackage[ansinew]{inputenc}
	\usepackage[usenames,dvipsnames]{pstricks}
    \usepackage{graphicx}
    \usepackage{float}
    \usepackage{subfigure}
	\usepackage{pst-grad} 
	\usepackage{pst-plot} 
	\usepackage{sidecap}
	\usepackage[makeroom]{cancel}

\begin{document}

\title[Distributing entangled state...]{Distributing entangled state using quantum repeater protocol: Trapped atomic ions in optomechanical cavities}

\author{M Ghasemi$^{1}$}

\author{M K Tavassoly$^{1}$}
\address{$^1$Atomic and Molecular Group, Faculty of Physics, Yazd University, Yazd  89195-741, Iran}
\ead{mktavassoly@yazd.ac.ir}

\vspace{10pt}
\date{today}

\begin{abstract}
Distribution of the entangled state of trapped atomic ions to long distance using quantum repeater protocol is considered. Indeed, the long distance is divided into short parts, and then using entanglement generation and entanglement swapping techniques in optomechanical cavities, the entanglement is distributed. To do the task, we perform interaction between trapped atomic ions in optomechanical cavities, operate proper measurements on trapped ions and also make Bell state measurement as a well-known way to swap the entanglement. Accordingly, the entanglement is distributed between target ions with satisfactory values of success probability and entanglement degree. The effects of detuning and amplitude of pump laser on the entanglement and success probability are evaluated. The fluctuations of entanglement and success probability are decreased by increasing of detuning. Via increasing the amplitude of pump laser, the maxima of entanglement are repeated more times and success probability undergoes the collapse-revival phenomenon.
\end{abstract}


\vspace{2pc}

\section{Introduction}
\begin{figure}[H]
            \centering
          \includegraphics[width=0.65\textwidth]{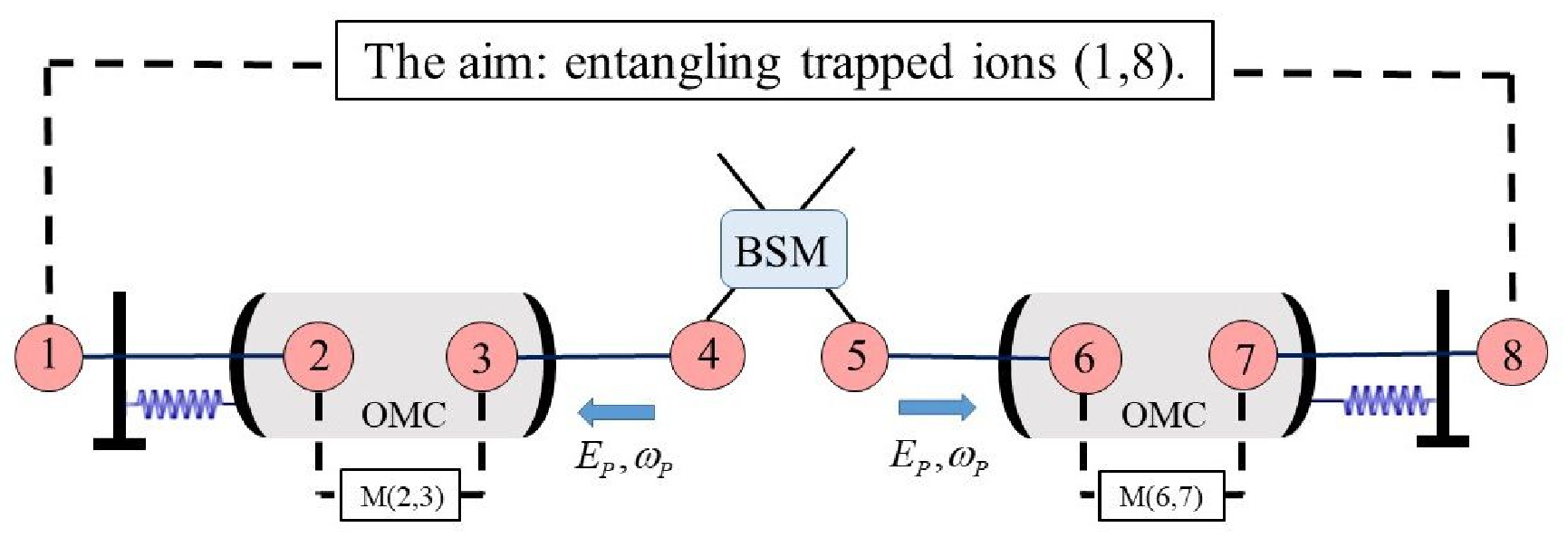}
            \caption{\label{fig.Fig1a} {The proposed quantum repeater protocol to distribute entangled state of trapped ions using the OMCs. Initially, we suppose that the ionic pairs $(i,i+1)$ with $i=1,3,5,7$, have been prepared in Bell-like states. After performing interaction between trapped atomic ions (2,3) and (6,7) in two independent OMCs and operating proper measurements on pairs (2,3) and (6,7) denoted by $\mathrm{M}(j,j+1)$ where $j=2,6$, the entanglement is generated between trapped ions (1,4) and (5,8), respectively. Finally, the target pair (1,8) is converted to the entangled state using BSM method shown in figure.}}
           \end{figure}

The distribution of quantum states is limited because of the existence of noise in optical channels.
  Quantum repeater protocol, which  plays a key role in quantum information transformation, especially for long distance applications, first introduced by Briegel et al \cite{Briegel1998}.
  In practice, the distributed entangled state between two points with long distance can be used as a quantum entangled channel for quantum teleportation purposes.
 In our quantum repeater protocol, the long distance between two target locations is divided into several short parts.
  Each of these parts is occupied distinctly by an entangled state, while each entangled pair is separable from all
   other pairs, even from the adjacent pair(s).
  Then, performing the entanglement swapping processes appropriately (to the separable pairs sequentially) the entanglement  is finally distributed between the two end locations \cite{Chen2007}. Therefore, there exist two key points herein: initial entangled pairs and entanglement swapping process. At first, the entanglement can be generally generated by performing interaction \cite{Zhao2012,Ye2004,Ye2007,Zou2003,Dalafi2019,Chakraborty2018,Li2015} or using beam splitter \cite{Ghasemi2019,Agarwal2012}. Secondly, this generated entanglement can be swapped (to the separable pairs) using Bell state measurement (BSM) \cite{Ghasemi2017,Nourmandipour2016,Ghasemi2016}, beam splitter \cite{Pakniat2017} or via performing interaction between independent particles (QED method) \cite{Pakniat22017}.
  In quantum repeater protocols, after producing entanglement and swapping entanglement between separable parts appropriately, these entangled states are stored in quantum memories. In quantum memories these entangled states are purified and so are released at an appropriate time. Moreover, the distribution of entangled states using quantum repeater protocols without quantum memories is also possible. In this regard, recently an experimental quantum repeater has been designed and analyzed without quantum memory \cite{Li2019}.\\
    Quantum repeater has been investigated in different systems such as  quantum dots \cite{Li2016,Yi2019}, trapped ions \cite{Sangouard2009}, Rydberg gates \cite{Zhao2010} and photonic system \cite{Azuma2015}. We have already considered quantum repeater protocols based on atomic systems in the presence of optomechanical cavity (OMC) \cite{Ghasemi42019} as well as optical cavity in the absence \cite{Ghasemi2018} and presence of dissipation \cite{Ghasemi22019,Ghasemi32019}. Also, the distribution of entangled coherent states between two locations using quantum repeater protocol has been investigated \cite{Ghasemi2019}.\\
  In this paper, we consider the possibility of distribution of entangled state between trapped atomic ions using OMCs. Distinctly, the main advantage of ions,
   with respect to neutral atoms and also photons,
    is that ions can effectively be trapped, cooled and also are more controllable, therefore, they can be stored at arbitrary desired fixed positions in
     space \cite{Porras2004,Duan2003}. In addition, the internal states of ionic systems can be precisely manipulated using laser light,
     and can be measured with basically $100\%$ efficiency \cite{Leibfried2003}. Due to the mentioned  advantages,
     various studies on ions have been done; for instance, one may refer to the generation of entanglement with hundreds
      of trapped ions \cite{Bohnet2016}, designing quantum computer with trapped atomic ions \cite{Brown2016}
       and building a quantum simulator using trapped ions \cite{Marciniak2018} (for more information
        see also \cite{Yazdanpanah22017,Nadiki2018,Zheng2003,Vogel1995}).
  And about using the OMCs in our paper we preliminary recall that, entanglement generation in the presence of OMCs is, in general, an attractive topic (see \cite{Tian2013,Maimaiti2015,Deng2016}).
    Generally, this issue plays a key role in quantum teleportation, entanglement swapping and quantum repeater schemes such as our model.
    Since according to our scheme, we should have initially various adjacent entangled pairs.
    In this regard, for example, in \cite{Tian2013} the authors have considered an optomechanical interface for the generation of
    photon entanglement. The entanglement between two cavity output modes has been
    concentrated by measuring the phonon number of the mechanical mode \cite{Maimaiti2015}. Indeed, the development
    of OMCs has opened up a new possibility to entangle photons with different frequencies
    \cite{Maimaiti2015}. Optimizing the output-photon entanglement in multimode OMCs has
    been considered in \cite{Deng2016}.\\ Furthermore, using the OMCs, instead of optical cavities, has some noticeable  advantages, too. One of the principal advantages of optomechanical systems is the built in readout of mechanical motion via the light field transmitted through (or reflected from) the cavity \cite{Aspelmeyer2014}. Also, the optomechanical systems are among the best candidates for the observation of quantum phenomena in macroscopic systems \cite{Vanner2015}. In this respect, the quantum decoherence in macroscopic mechanical objects has been frequently studied using OMCs \cite{Romero2011,Marshall2003}. As another remarkable fact, one may refer to the range of OMCs which is from nanometer-sized devices to micromechanical
  structures and even to macroscopic centimeter-sized mirrors \cite{Aspelmeyer2012}. Therefore, as mentioned before, observing quantum phenomena and quantum technologies
  using some macroscopic systems is possible; this is so another interesting feature of using the OMCs
  in our quantum repeater model (in other words the quantum vibrational mode of OMCs which is their particular characteristics may be produced in macroscopic scales). In addition, due to the very long coherence times of the mechanical systems, the OMCs are suitable systems for the purpose of optical memories \cite{Xiao2014,Lvovsky2009}.
  In this respect, even though in our scheme we have not used optical memory (recall that designing the quantum repeater without quantum memory is also possible \cite{Li2019}), moreover, considering the above characteristics of OMCs in addition to this latter property propose the authors who are interested  in
  performing quantum repeaters containing quantum memories to use OMCs in their further schemes.\\
  Keeping in mind the above-mentioned advantages of ionic systems as well as the OMCs, we motivated to study distributing entangled state of trapped ions by performing
  interaction between the trapped ions in OMCs in our
  introduced quantum repeater protocol (see figure \ref{fig.Fig1a}). This protocol includes eight two-level trapped atomic ions labeled by $(1,2,\cdots, 8)$ where the pairs $(i,i+1)$ with $i=1,3,5,7$ have been initially prepared in maximally entangled states. As shown in this figure, we first perform entanglement swapping via
              appropriate interaction between trapped ions (2,3) and (6,7) in two independent OMCs.
              Then, by performing proper measurements (denoted by $\mathrm{M}(j,j+1)$ where $j=2,6$, in figure \ref{fig.Fig1a}) on trapped ions (2,3) and (6,7), \textit{i.e.,} which is done by operating appropriate projection operators, the
              separable trapped ions (1,4) and (5,8) are respectively converted to the entangled states.
              %
                        Anyway, these states can be entangled or separable depending
                            on the time. The success probabilities and concurrences of these states are calculated. At last, by performing the projective measurement with the proper Bell state which is constructed by trapped ions (4,5) on the total state of trapped ions (1,4,5,8), the BSM method is implemented and the entanglement
              is suitably swapped to target trapped ions (1,8). The success probability and concurrence of this target state are numerically calculated.
              Our quantum repeater protocol is studied in the presence of high-Q cavities and enough low temperature (\textit{i.e.,} in the absence of dissipation effects). The dissipation effects essentially exist in all systems, however, it can be reduced via adjusting the involved parameters in the considered model. To achieve the condition, when detuning $\Delta$ is enough large, \textit{i.e.,} $\Delta >> \gamma$,
      the spontaneous decay rate $\gamma$ from atomic levels can be neglected \cite{Cao2006}. Under the condition that the trap frequency is much greater than the atomic decay rate, it is possible to ignore the effects of atomic decay \cite{Wu1997}. In addition, the effect of cavity losses on the system may also be neglected if the Q
      factor of the cavity is high enough \cite{Nasreen1993} and the thermal noise of mirrors in OMC can be reduced by cold damping \cite{Courty2001}.\\
  \\
     This paper organizes as follows: Our model for quantum repeater protocol is introduced in Sec.
     \ref{model}. The numerical results which establish the suitability of our protocol are investigated in Sec. \ref{sec.results}.  Finally, our paper ends with a summary and conclusion in Sec. \ref{Summary}.
 \section{Quantum repeater model}\label{model}
   \begin{figure}[h]
   \centering
        \subfigure[\label{fig.Fig1b} \ The diagram of OMC]{\includegraphics[width=0.30\textwidth]{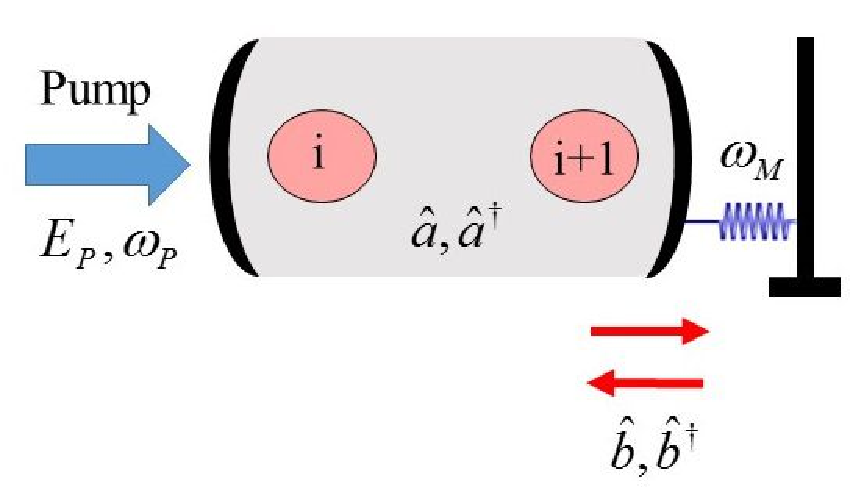}}
      \hspace{0.05\textwidth}
      \subfigure[\label{fig.Fig1c} \ The energy levels of trapped atomic ions]{\includegraphics[width=0.30\textwidth]{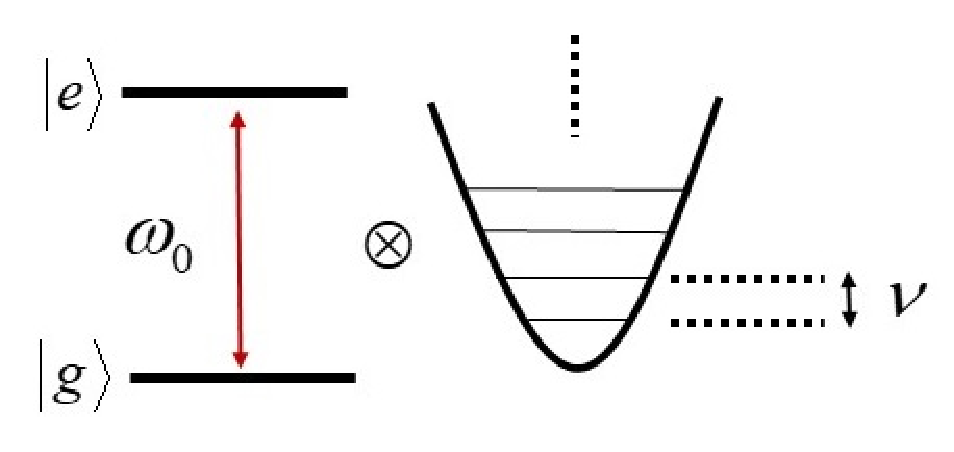}}
    \caption{\label{fig:Fig1} {(a) The OMC with one movable mirror contains pair $(i,i+1)$ where $i=2,6$. (b) The scheme of trapped atomic ion with ground (exited) state $\ket{g}$ ($\ket{e}$) and its vibrational modes.}}
   \end{figure}
 The initial state of trapped ions $(1,2,3,4)$ of our quantum repeater model described in the Introduction section reads as
 \begin{eqnarray}\label{initialstate}
 \ket{\psi}_{1-4}&=&\ket{\psi}_{1,2}\otimes \ket{\psi}_{3,4},
   \end{eqnarray}
 where the  pairs $(1,2)$ and $(3,4)$ are supposed to be prepared in the Bell-like states as below:
 \begin{eqnarray}\label{initialstates}
 \ket{\psi}_{i,i+1}&=&\frac{1}{\sqrt{2}}(\ket{\cal E}_i\ket{\cal G}_{i+1}+\ket{\cal G}_i\ket{\cal E}_{i+1}), \qquad   i=1,3,
   \end{eqnarray}
  with the definitions $\ket{\cal E}\equiv\ket{e,1}$ and $\ket{\cal G}\equiv\ket{g,0}$, where $\ket{e}$ and $\ket{g}$ are the ion's internal levels and $\ket{1}$ and $\ket{0}$ are the number states of the vibrational mode of the ions (see figure \ref{fig.Fig1c}).

  Even though various entangled internal states and entangled states of vibrational modes typically nearly to our considered entangled state have been previously generated between two trapped atomic ions using different methods, we ourselves tried to produce the necessary entangled state in our protocol in Eq. (\ref{initialstates})
  (see Appendix A).
    For instance, the combination of a purely dispersive with a resonant laser excitation of vibronic
               transitions of the ions leads to the generation of internal Bell states of two trapped
               ions \cite{Solano1999}. Also, the entangled internal states can be created between two trapped atomic ions by an
               interference effect and state projection accompanying a measurement \cite{Blatt2008}. In Ref. \cite{Munro2000}, a scheme has been proposed for entanglement generation between two
               vibrational modes of two ions. Then, to swap the entanglement to trapped atomic ions (1,4), two separable trapped atomic ions $(2,3)$ are sent to an OMC with one movable mirror (see figure \ref{fig.Fig1b}), while the initial states of optical and mechanical modes are supposed to be in their vacuum states. In this regard, the interaction between a two-level trapped atomic ion inside a single-mode
                                  optical cavity in the Lamb-Dick regime
                                  as well as the first vibrational sideband in the rotating wave approximation has been previously
                                  introduced \cite{Buzek1997,Yazdanpanah2016}. Also, a system including a two-level trapped atomic ion interacting with an OMC has been studied in \cite{Nadiki2018}.\\ The interaction between two trapped atomic ions (2,3) in an OMC is described by the Hamiltonian $\hat{H}=\hat{H}_0+\hat{H}_1$ $(\hbar=1)$, where
 \begin{eqnarray}\label{hamiltonian}
  \hat{H}_0&=&\omega_c\hat{a}^{\dagger}\hat{a}+\omega_M\hat{b}^{\dagger}\hat{b}+\sum_{i=2,3}\left(\nu \hat{c}_i^{\dagger}\hat{c}_i+\dfrac{\omega_0}{2}\hat{\sigma}_z^{(i)} \right) ,\\\nonumber
  \hat{H}_1&=&-G\hat{a}^{\dagger}\hat{a}(\hat{b}+\hat{b}^{\dagger})+i\sum_{i=2,3} g_i(\hat{a}\hat{\Sigma}_{+}^{(i)}-\hat{a}^{\dagger}\hat{\Sigma}_{-}^{(i)})  -i E_P  \left(\hat{a} e^{i \omega_P t}-\hat{a}^{\dagger} e^{-i \omega_P t}\right).
   \end{eqnarray}
   In this Hamiltonian $\omega_c$ ($\omega_M$) is the frequency of optical (mechanical) mode with $\hat{a}$, $\hat{a}^{\dagger}$ ($\hat{b}$, $\hat{b}^{\dagger}$) as the annihilation and creation operators associated with optical (vibrational) mode. Also, the third term in $\hat{H}_0$ corresponds to the trapped atomic ions $(2,3)$, where $\nu$ is the frequency of the vibrational mode of $i$th ion with $\hat{c}_i$, $\hat{c}^{\dagger}_i$ as its annihilation, creation operators. $\hat{\sigma}_z^{(i)}$ is the ionic population inversion (the atomic levels of trapped ions have been shown in figure \ref{fig.Fig1c}), wherein the frequency of ionic transition has been indicated by $\omega_0$. In the Hamiltonian $\hat{H}_1$ of Eq. (\ref{hamiltonian}), $G$ and $g_i$ are respectively the optomechanical coupling strength to the field mode and coupling constant between $i$th trapped ion and optical mode. The operator $\hat{\Sigma}_{-}^{(i)}=\hat{c}\hat{\sigma}_{-}^{(i)}$ ($\hat{\Sigma}_{+}^{(i)}=\hat{c}^{\dagger}\hat{\sigma}_{+}^{(i)}$) where $\sigma_{-}^{(i)}=\ket{g}^{(i)}\bra{e}(\sigma_{+}^{(i)}=\sigma_{-}^{(i)\dagger})$ is the lowering (raising) operator of $i$th trapped ion. Also, we suppose that our system is driven by a pump laser with frequency $\omega_P$ and amplitude $E_P$.
       Now, we can obtain the effective Hamiltonian of the system with the help of proposed approach in \cite{James2007,Gamel2010} as below (for more detail see Appendix B):
       \begingroup\makeatletter\def\f@size{6}\check@mathfonts
  \begin{eqnarray}\label{eff}
    \hat{H}^\mathrm{eff}_{(2,3)}&=&-\dfrac{g^2_2}{\omega_{22}}\hat{\Sigma}_{+}^{(2)}\hat{\Sigma}_{-}^{(2)}-\dfrac{g^2_3}{\omega_{33}}\hat{\Sigma}_{+}^{(3)}\hat{\Sigma}_{-}^{(3)}-\dfrac{g_2 g_3}{\omega_{23}}\left( \hat{\Sigma}_{-}^{(2)}\hat{\Sigma}_{+}^{(3)}e^{i(\omega_2-\omega_3)t}+\hat{\Sigma}_{-}^{(3)}\hat{\Sigma}_{+}^{(2)}e^{-i(\omega_2-\omega_3)t}\right)\\\nonumber
    &+&\dfrac{g_2 E_P}{\omega_{24}}\left( \hat{\Sigma}_{+}^{(2)}e^{i(\omega_4-\omega_2)t}+\hat{\Sigma}_{-}^{(2)}e^{-i(\omega_4-\omega_2)t}\right)+\dfrac{g_3 E_P}{\omega_{34}}\left( \hat{\Sigma}_{+}^{(3)}e^{i(\omega_4-\omega_3)t}+\hat{\Sigma}_{-}^{(3)}e^{-i(\omega_4-\omega_3)t}\right),
        \end{eqnarray}
 where the detuning $\omega_n$ and $\omega_{ij}$ with $n,i,j=1,2,3,4$ have been introduced in Appendix B. Henceforth, the time-dependent state of trapped ions (1,2,3,4) may be obtained with the help of initial trapped ionic state (\ref{initialstate}), the effective Hamiltonian (\ref{eff}) and time-dependent Schr\"{o}dinger equation ?$i\frac{\partial }{\partial t}\ket{\psi(t)}=?\hat{H}^\mathrm{eff}_{(2,3)}\ket{\psi(t)}$? as below (see Appendix C for details of the calculation of the  coefficients of this state):
 \begin{eqnarray}\label{state1-4}
 \ket{\psi(t)}=\sum^4_{i=1} \ket{\Phi}^i_{2,3}\otimes\ket{\Psi(t)}^i_{1,4},
 \end{eqnarray}
 where $\ket{\Phi}^1_{2,3}=\ket{\cal G}_2\ket{\cal E}_3$, $\ket{\Phi}^2_{2,3}=\ket{\cal E}_2\ket{\cal G}_3$, $\ket{\Phi}^3_{2,3}=\ket{\cal E}_2\ket{\cal E}_3$ and $\ket{\Phi}^4_{2,3}=\ket{\cal G}_2\ket{\cal G}_3$. Also, for simplicity the states of trapped ions (1,4) identified in Eq. (\ref{state1-4}) can be expressed as follow,
 \begin{eqnarray}\label{coef0}
 \ket{\Psi(t)}^i_{1,4}&=&\left(\alpha_i(t) \ket{\cal E}_1\ket{\cal G}_4+\beta_i(t) \ket{\cal G}_1\ket{\cal E}_4\right. +\left. \gamma_i(t) \ket{\cal G}_1\ket{\cal G}_4+\eta_i(t) \ket{\cal E}_1\ket{\cal E}_4\right),
 \end{eqnarray}
with $i=1,2,3,4$.
  Paying attention to the state (\ref{state1-4}), there are four time-dependent states $\ket{\Psi(t)}^i_{1,4}$ for trapped ions $(1,4)$ after applying a projective measurement with the state $\ket{\Phi}^i_{2,3}$ on state (\ref{state1-4}). The states $\ket{\Psi(t)}^i_{1,4}$ can be entangled or separable depending
        on the time. The success probability and concurrence \cite{Romero2007} for the obtained state $\ket{\Psi(t)}^i_{1,4}$ introduced in Eq. (\ref{state1-4}) after applying a projective measurement are respectively calculated as,
  \begin{eqnarray}\label{ent14}
   P^i_{1,4}(t)&=&\left| \alpha_i(t)\right|^2+\left| \beta_i(t)\right|^2+\left| \gamma_i(t)\right|^2+\left| \eta_i(t)\right|^2,\\ \nonumber
      C^i_{1,4}(t)&=&\dfrac{2}{ P^i_{1,4}(t)}\left|\eta_i(t) \gamma_i(t)-\alpha_i(t) \beta_i(t) \right|.
    \end{eqnarray}
 The above processes can straightforwardly be repeated for ions $(5-8)$ and the time-dependent states $\ket{\Psi(t)}^i_{5,8}$ where $\ket{\Psi(t)}^i_{5,8}=\ket{\Psi(t)}^i_{1,4}$ are generated for ions $(5,8)$ with success probability $P^i_{5,8}(t)=P^i_{1,4}(t)$ and concurrence $C^i_{5,8}(t)=C^i_{1,4}(t)$.
 Clearly, the total state of ions labeled by $(1,4,5,8)$ reads as,
 \begin{eqnarray}\label{state1458}
 \ket{\Psi(t)}^{i,j}_{1,4,5,8}&=&\ket{\widetilde{\Psi(t)}}^i_{1,4}\otimes\ket{\widetilde{\Psi(t)}}^j_{5,8},\qquad i,j=1,2,3,4,
  \end{eqnarray}
 where $\ket{\widetilde{\Psi(t)}}^{i}_{1,4(5,8)}$ is the normalized of state $\ket{\Psi(t)}^{i}_{1,4(5,8)}$. Now, the generated entanglement between trapped ions (1, 4) and (5, 8) is swapped to target trapped ions (1, 8) by performing the projective measurement $\ket{\psi}_{4,5}\bra{\psi}$ with the Bell-like state $\ket{\psi}_{4,5}=\dfrac{1}{\sqrt{2}}\left(\ket{\cal E}_4\ket{\cal E}_5+\ket{\cal G}_4\ket{\cal G}_5 \right) $ on state (\ref{state1458}) \cite{Liao2011}. After this measurement, the state of ions $(1,8)$ is obtained as below:
  \begin{eqnarray}\label{state118}
 \ket{\Psi(t)}^{i,j}_{1,8} &=&\dfrac{1}{\sqrt{N_{i,j}(t)}}\left[ \left(\alpha_i(t)\gamma_j(t)+\eta_i(t) \alpha_j(t)\right)\ket{\cal E}_1\ket{\cal G}_8\right. \\\nonumber
  &+&\left. \left(\gamma_i(t)\beta_j(t)+\beta_i(t) \eta_j(t)\right)\ket{\cal G}_1\ket{\cal E}_8\right. +\left. \left(\gamma_i(t)\gamma_j(t)+\beta_i(t) \alpha_j(t)\right)\ket{\cal G}_1\ket{\cal G}_8\right. \\\nonumber
   &+&\left. \left(\alpha_i(t)\beta_j(t)+\eta_i(t) \eta_j(t)\right)\ket{\cal E}_1\ket{\cal E}_8\right],\\\nonumber
 N_{i,j}(t)&=& \left|\alpha_i(t)\gamma_j(t)+\eta_i(t) \alpha_j(t) \right|^2+\left|\gamma_i(t)\beta_j(t)+\beta_i(t) \eta_j(t) \right| ^2\\\nonumber
 &+&\left|\gamma_i(t)\gamma_j(t)+\beta_i(t) \alpha_j(t) \right| ^2+\left|\alpha_i(t)\beta_j(t)+\eta_i(t) \eta_j(t) \right| ^2.
   \end{eqnarray}
   The success probability and concurrence for the time-dependent state (\ref{state118}) are respectively calculated as,
   \begin{eqnarray}\label{ent18}
   \small
    P^{i,j}_{1,8}(t)&=&\dfrac{N_{i,j}(t)}{2 \left| P^i_{1,4}(t)\right| \left| P^j_{5,8}(t)\right| },\\ \nonumber
     C^{i,j}_{1,8}(t)&=&\dfrac{2}{N_{i,j}(t)}\left|\left(\alpha_i(t)\beta_j(t)+\eta_i(t) \eta_j(t)\right)\right. \left(\gamma_i(t)\gamma_j(t)+\beta_i(t) \alpha_j(t)\right)\\ \nonumber
       &-& \left(\alpha_i(t)\gamma_j(t)+\eta_i(t) \alpha_j(t)\right)\left. \left(\gamma_i(t)\beta_j(t)+\beta_i(t) \eta_j(t)\right) \right| .
     \end{eqnarray}
  Similarly, the other state of ions $(1,8)$ may be obtained by performing the projective measurement $\ket{\psi^{'}}_{4,5}\bra{\psi^{'}}$ with the Bell-like state $\ket{\psi^{'}}_{4,5}=\dfrac{1}{\sqrt{2}}\left(\ket{\cal E}_4\ket{\cal G}_5+\ket{\cal G}_4\ket{\cal E}_5 \right) $ on the state (\ref{state1458}) as below:
  \begin{eqnarray}\label{state218}
 \ket{\Psi^{'}(t)}^{i,j}_{1,8}&=&\dfrac{1}{\sqrt{N^{'}_{i,j}(t)}}\left[ \left(\alpha_i(t)\alpha_j(t)+\eta_i(t) \gamma_j(t)\right)\ket{\cal E}_1\ket{\cal G}_8\right. \\\nonumber
  &+&\left. \left(\gamma_i(t)\eta_j(t)+\beta_i(t) \beta_j(t)\right)\ket{\cal G}_1\ket{\cal E}_8\right. +\left. \left(\gamma_i(t)\alpha_j(t)+\beta_i(t) \gamma_j(t)\right)\ket{\cal G}_1\ket{\cal G}_8\right. \\\nonumber
  &+&\left. \left(\alpha_i(t)\eta_j(t)+\eta_i(t) \beta_j(t)\right)\ket{\cal E}_1\ket{\cal E}_8\right],\\\nonumber
  N^{'}_{i,j}(t)&=& \left|\alpha_i(t)\alpha_j(t)+\eta_i(t) \gamma_j(t) \right|^2+\left|\gamma_i(t)\eta_j(t)+\beta_i(t) \beta_j(t) \right|^2 \\\nonumber
     &+& \left|\gamma_i(t)\alpha_j(t)+\beta_i(t) \gamma_j(t) \right|^2+\left|\alpha_i(t)\eta_j(t)+\eta_i(t) \beta_j(t) \right|^2.
   \end{eqnarray}
   The success probability and concurrence for state (\ref{state218}) are respectively calculated as
   \begin{eqnarray}\label{ent218}
   \small
    P^{'i,j}_{1,8}(t)&=&\dfrac{ N^{'}_{i,j}(t)}{2 \left| P^i_{1,4}(t)\right| \left| P^j_{5,8}(t)\right| },\\ \nonumber
       C^{'i,j}_{1,8}(t)&=&\dfrac{2}{N^{'}_{i,j}(t)}\left| \left(\alpha_i(t)\eta_j(t)+\eta_i(t) \beta_j(t)\right)\right. \left(\gamma_i(t)\alpha_j(t)+\beta_i(t) \gamma_j(t)\right)\\ \nonumber
       &-& \left(\alpha_i(t)\alpha_j(t)+\eta_i(t) \gamma_j(t)\right)\left. \left(\gamma_i(t)\eta_j(t)+\beta_i(t) \beta_j(t)\right) \right| .
     \end{eqnarray}
 In the next section, to reach an intuitive sense about the above quantum repeater model, our numerical results for the important quantities which are calculated in above, \textit{i.e.,} concurrences and success probabilities, are evaluated and discussed.
  \section{Results and discussion} \label{sec.results}
 In this section, the entanglement (concurrence) and success probabilities of distributed time-dependent states of trapped ions $(1,4)$,  $(5,8)$ and $(1,8)$ are evaluated through which the effects of detuning, $\omega_n$, $n=1,2,3,4$, and amplitude of pump laser, $E_P$, are studied. These analysis are discussed for Eqs. (\ref{ent14}), (\ref{ent18}) and (\ref{ent218}) with $i=j=1$ by considering $\omega_4=\omega_M$ and $\nu+\omega_0=\omega_P$, \textit{i.e.,} the pump laser is always red detuned and $\omega_1=\omega_2=\omega_3=\omega_4=\omega_M$. The chaotic and intense fluctuations of concurrence of trapped ions $(1,4)$ can be seen (figure \ref{fig.Fig2a}) with acceptable stable success probability (figure \ref{fig.Fig2b}). Also, in figures \ref{fig.Fig2c} and \ref{fig.Fig2e}, the fluctuations of concurrence of entangled states for trapped ions $(1,8)$ obtained via performing different projective measurements can be observed. From figures \ref{fig.Fig2d} and \ref{fig.Fig2f}, one can observe that the maxima of success probability for these states have been reached to the acceptable value of $0.5$ \cite{Ralph2001,Knill2001,Bergou2005}.\\
 The effect of detuning has been considered in figure \ref{fig.fig3}. As is seen from figure \ref{fig.Fig3a}, the intensity of fluctuations of entanglement between trapped ions $(1,4)$ has been decreased by increasing the amount of detuning. But, as shown in figure \ref{fig.Fig3b} success probability of the generated entangled state of trapped ions $(1,4)$ is independent of detuning and time. In figures \ref{fig.Fig3c}, \ref{fig.Fig3e} (\ref{fig.Fig3d}, \ref{fig.Fig3f}) the intensity of fluctuations of entanglement between trapped ions $(1,8)$ (success probability of the distributed entangled state of ions $(1,8)$) has been decreased by increasing the amount of detuning.\\
 The effect of amplitude of pump laser has been considered in figure \ref{fig.fig4}. As is seen from figure \ref{fig.Fig4a}, the periodic behaviour of concurrence (entanglement between trapped ions $(1,4)$) can be revealed by increasing the amount of $E_P$. Also, one can see from figure \ref{fig.Fig4b} that, the success probability of entangled state of trapped ions (1,4) is independent of the amplitude of pump laser. In figures \ref{fig.Fig4c}, \ref{fig.Fig4e} the maxima of entanglement have been achieved in more times by increasing the amplitude of pump laser. Also, in figures \ref{fig.Fig4d} and \ref{fig.Fig4f}, the collapse-revival phenomenon has been appeared for success probability by increasing the amplitude of pump laser.
     \begin{figure}[H]
          \centering
          \subfigure[\label{fig.Fig2a} \ $C^1_{1,4}(t)$]{\includegraphics[width=0.30\textwidth]{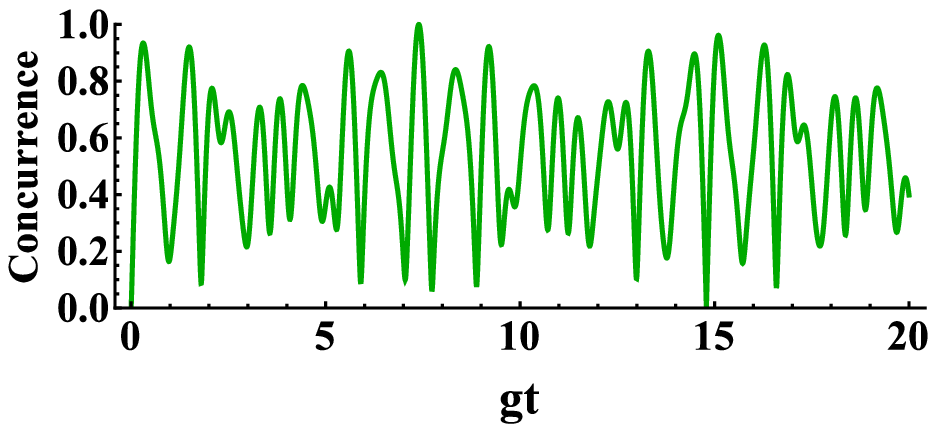}}
          \hspace{0.05\textwidth}
          \subfigure[\label{fig.Fig2b} \ $P^1_{1,4}(t)$]{\includegraphics[width=0.30\textwidth]{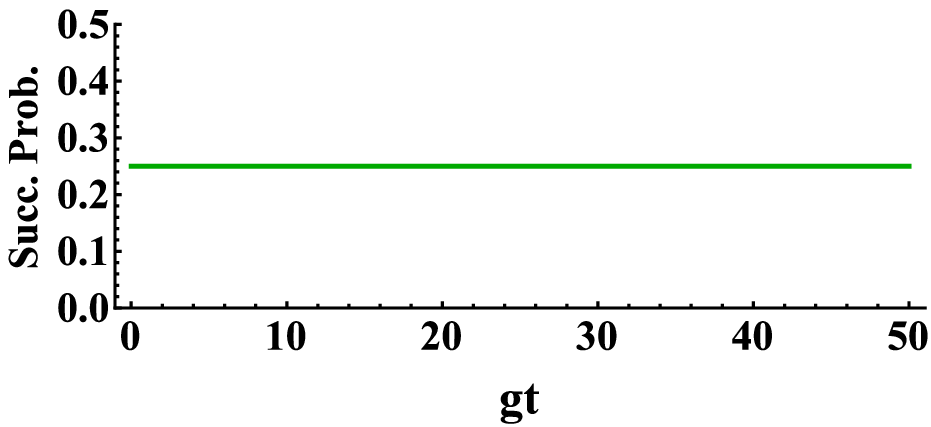}}
           \hspace{0.05\textwidth}
          \subfigure[\label{fig.Fig2c} \ $C^{1,1}_{1,8}(t)$]{\includegraphics[width=0.30\textwidth]{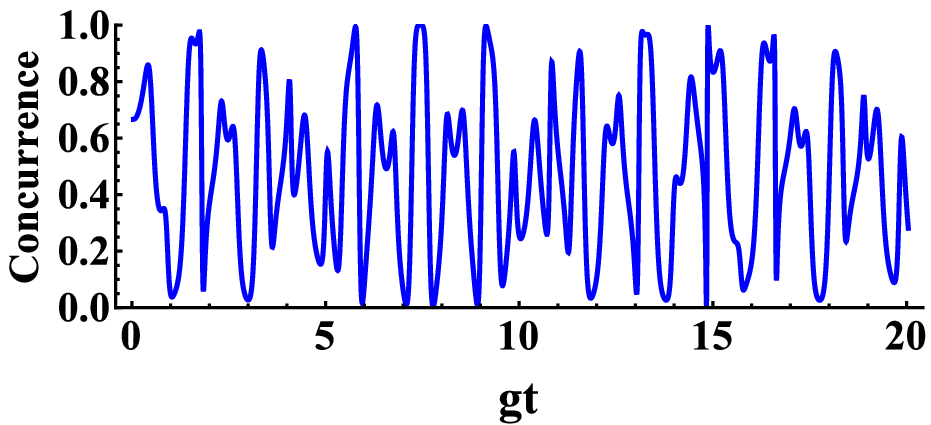}}
         \hspace{0.05\textwidth}
        \subfigure[\label{fig.Fig2d} \ $P^{1,1}_{1,8}(t)$]{\includegraphics[width=0.30\textwidth]{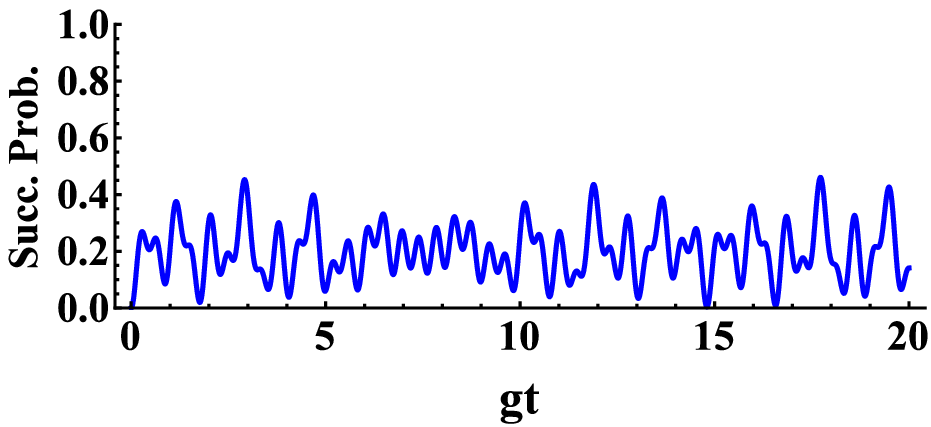}}
         \hspace{0.05\textwidth}
                 \subfigure[\label{fig.Fig2e} \ $C^{'1,1}_{1,8}(t)$]{\includegraphics[width=0.30\textwidth]{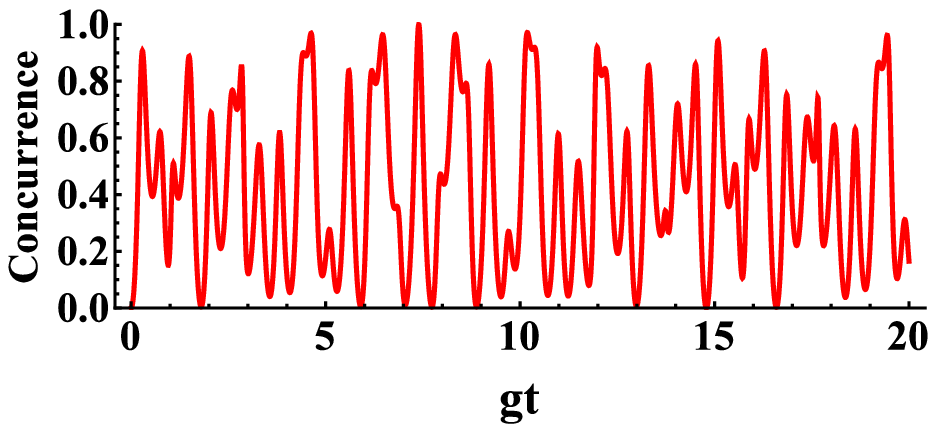}}
                \hspace{0.05\textwidth}
               \subfigure[\label{fig.Fig2f} \ $P^{'1,1}_{1,8}(t)$]{\includegraphics[width=0.30\textwidth]{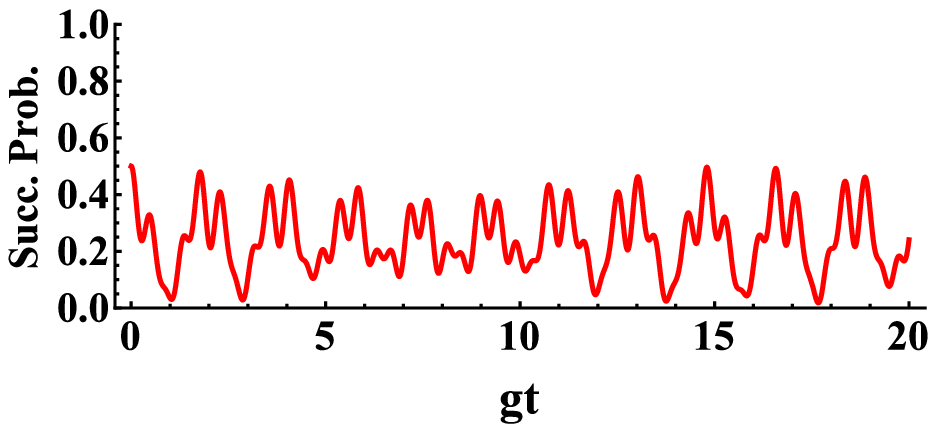}}
          \caption{\label{fig.fig2} {\it Evolution of} (a) concurrence $C^1_{1,4}(t)$ and (b) success probability $P^1_{1,4}(t)$ both related to Eq. (\ref{ent14}) (solid green line), (c) concurrence $C^{1,1}_{1,8}(t)$ and (d) success probability $P^{1,1}_{1,8}(t)$ both related to Eq. (\ref{ent18}) (solid blue line), (e) concurrence $C^{'1,1}_{1,8}(t)$ and (f) success probability $P^{'1,1}_{1,8}(t)$ both related to Eq. (\ref{ent218}) (solid red line) for $\omega_4=\omega_M=0.4g$, $E_P=0.5g$ with $g_2=g_3=g$, $\nu+\omega_0=\omega_P$, $gt>0$.}
          \end{figure}
     \begin{figure}[H]
           \centering
           \subfigure[\label{fig.Fig3a} \ $C^1_{1,4}(t)$]{\includegraphics[width=0.30\textwidth]{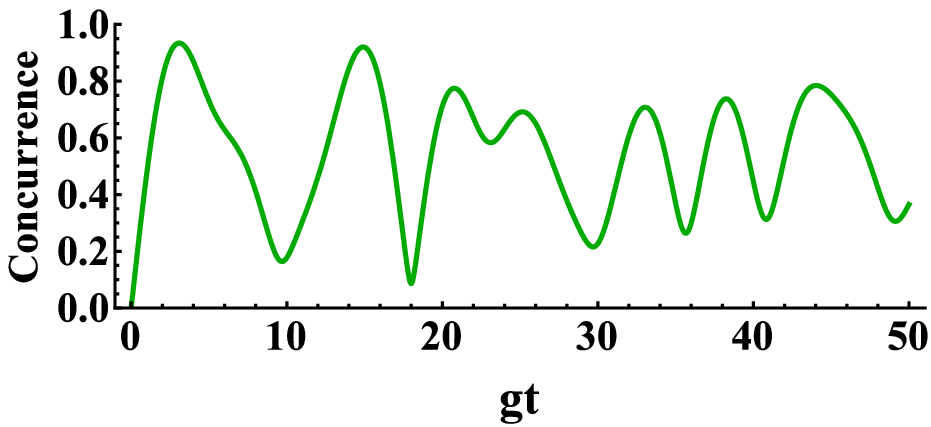}}
           \hspace{0.05\textwidth}
           \subfigure[\label{fig.Fig3b} \ $P^1_{1,4}(t)$]{\includegraphics[width=0.30\textwidth]{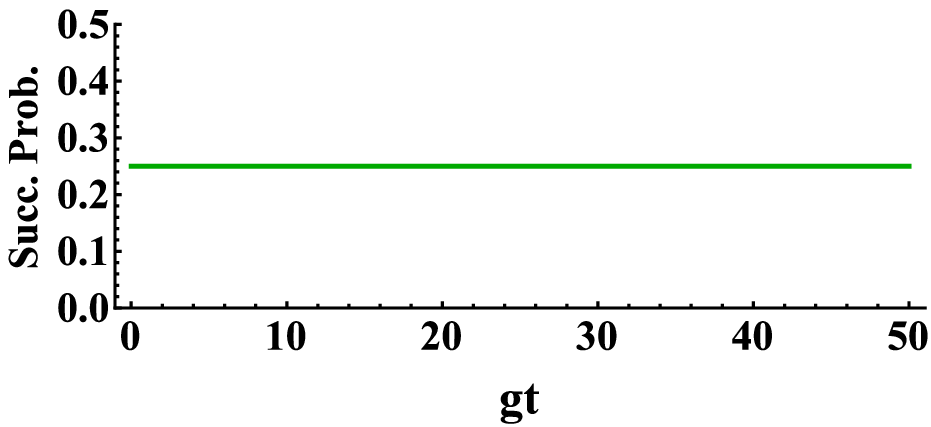}}
            \hspace{0.05\textwidth}
           \subfigure[\label{fig.Fig3c} \ $C^{1,1}_{1,8}(t)$]{\includegraphics[width=0.30\textwidth]{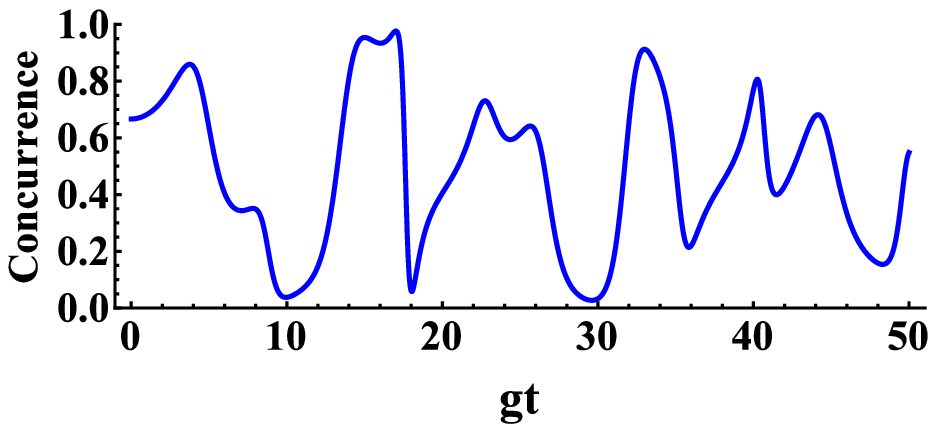}}
          \hspace{0.05\textwidth}
         \subfigure[\label{fig.Fig3d} \ $P^{1,1}_{1,8}(t)$]{\includegraphics[width=0.30\textwidth]{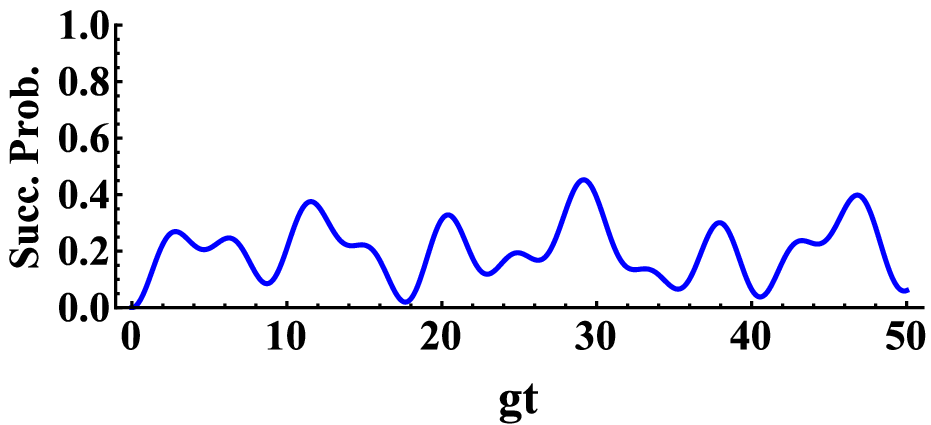}}
          \hspace{0.05\textwidth}
                   \subfigure[\label{fig.Fig3e} \ $C^{'1,1}_{1,8}(t)$]{\includegraphics[width=0.30\textwidth]{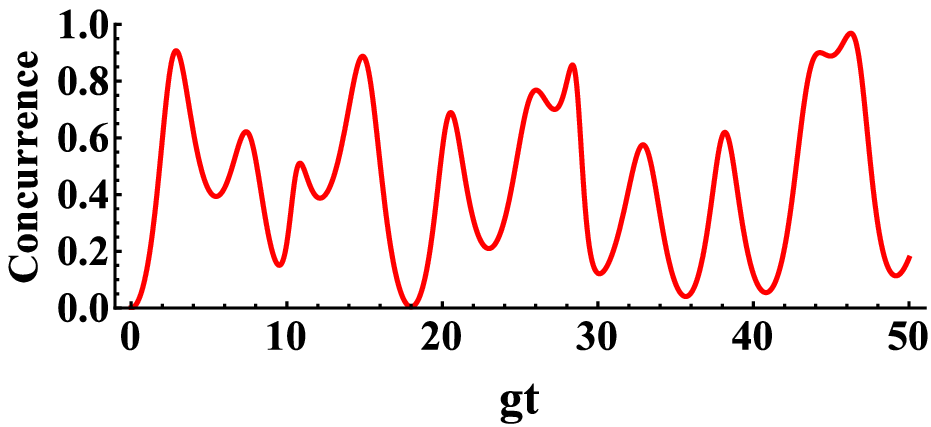}}
                  \hspace{0.05\textwidth}
                 \subfigure[\label{fig.Fig3f} \ $P^{'1,1}_{1,8}(t)$]{\includegraphics[width=0.30\textwidth]{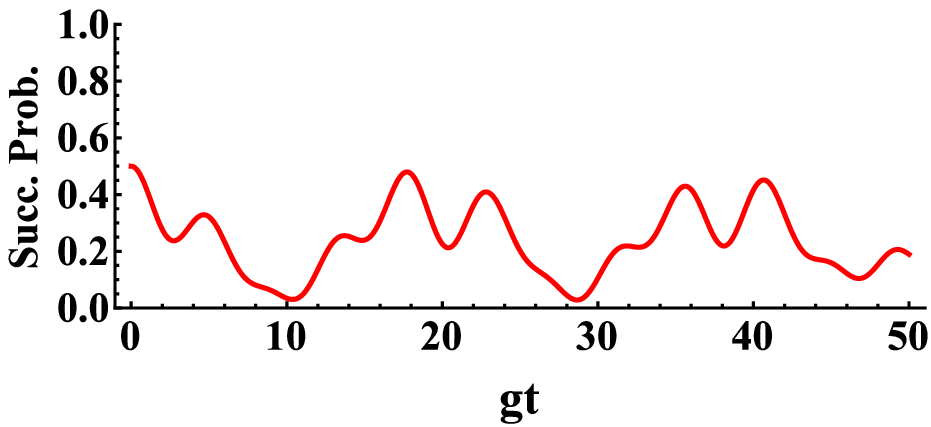}}
           \caption{\label{fig.fig3} {\it Evolution of} (a) concurrence $C^1_{1,4}(t)$ and (b) success probability $P^1_{1,4}(t)$ both related to Eq. (\ref{ent14}) (solid green line), (c) concurrence $C^{1,1}_{1,8}(t)$ and (d) success probability $P^{1,1}_{1,8}(t)$ both related to Eq. (\ref{ent18}) (solid blue line), (e) concurrence $C^{'1,1}_{1,8}(t)$ and (f) success probability $P^{'1,1}_{1,8}(t)$ both related to Eq. (\ref{ent218}) (solid red line) for $\omega_4=\omega_M=4g$, $E_P=0.5g$ with $g_2=g_3=g$, $\nu+\omega_0=\omega_P$, $gt>0$.}
           \end{figure}
      \begin{figure}[H]
               \centering
               \subfigure[\label{fig.Fig4a} \ $C^1_{1,4}(t)$]{\includegraphics[width=0.30\textwidth]{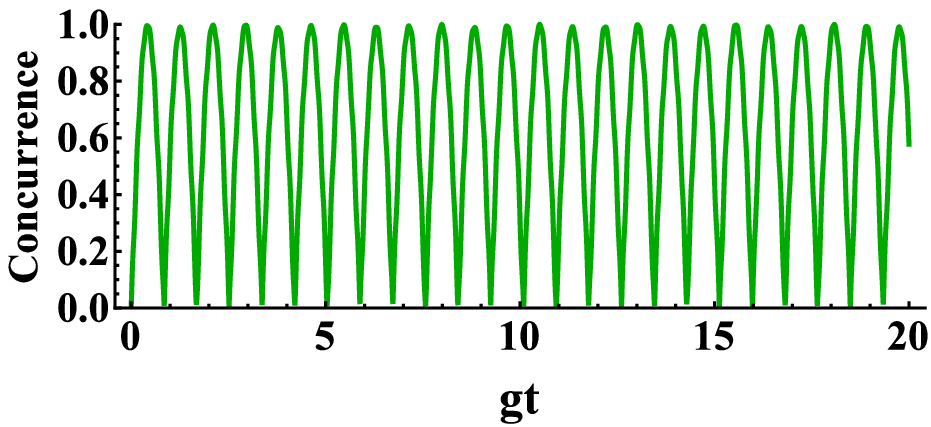}}
               \hspace{0.05\textwidth}
               \subfigure[\label{fig.Fig4b} \ $P^1_{1,4}(t)$]{\includegraphics[width=0.30\textwidth]{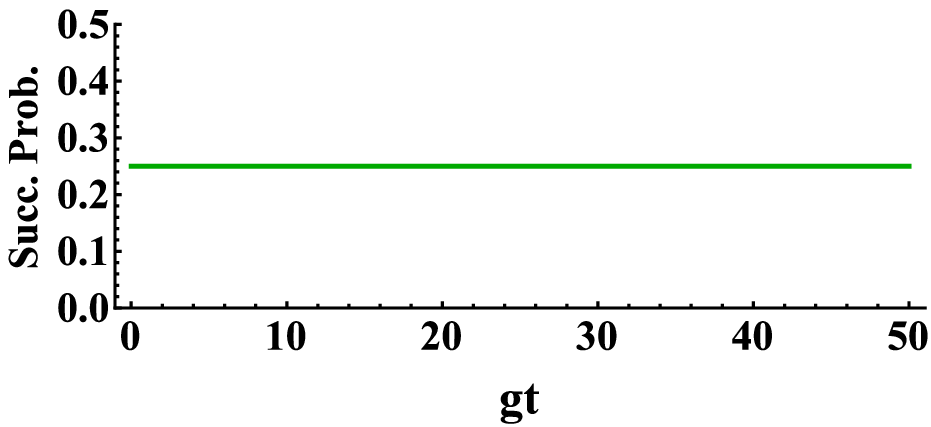}}
                \hspace{0.05\textwidth}
               \subfigure[\label{fig.Fig4c} \ $C^{1,1}_{1,8}(t)$]{\includegraphics[width=0.30\textwidth]{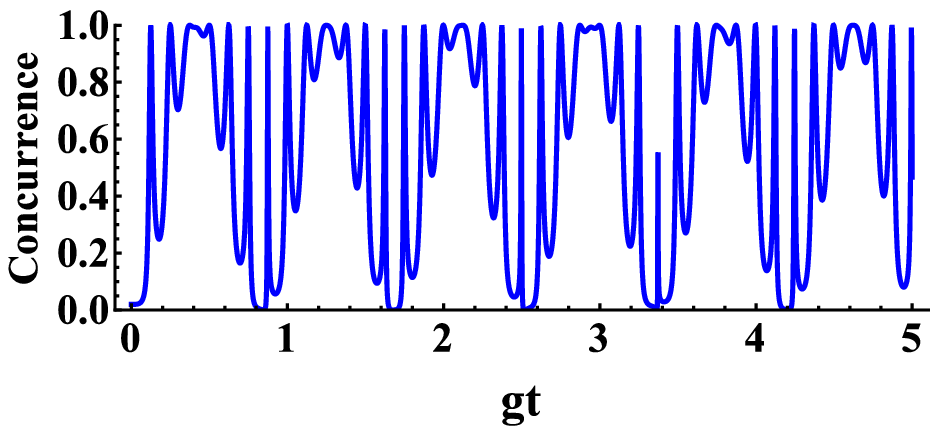}}
              \hspace{0.05\textwidth}
             \subfigure[\label{fig.Fig4d} \ $P^{1,1}_{1,8}(t)$]{\includegraphics[width=0.30\textwidth]{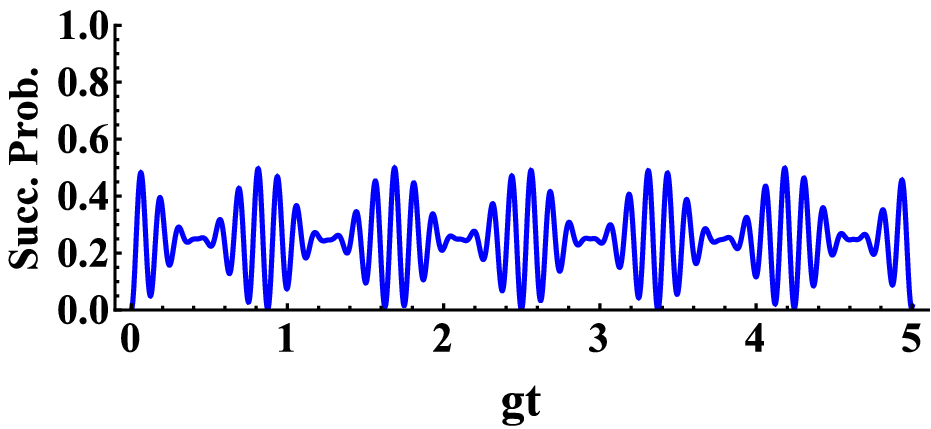}}
              \hspace{0.05\textwidth}
                           \subfigure[\label{fig.Fig4e} \ $C^{'1,1}_{1,8}(t)$]{\includegraphics[width=0.30\textwidth]{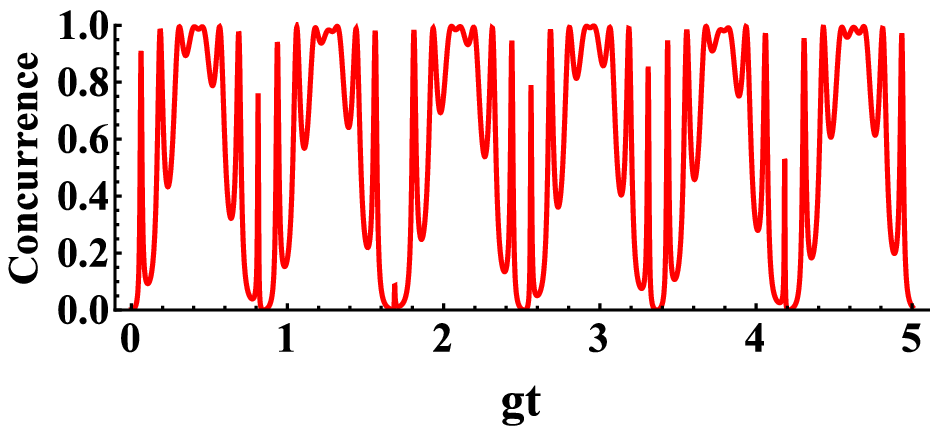}}
                          \hspace{0.05\textwidth}
                         \subfigure[\label{fig.Fig4f} \ $P^{'1,1}_{1,8}(t)$]{\includegraphics[width=0.30\textwidth]{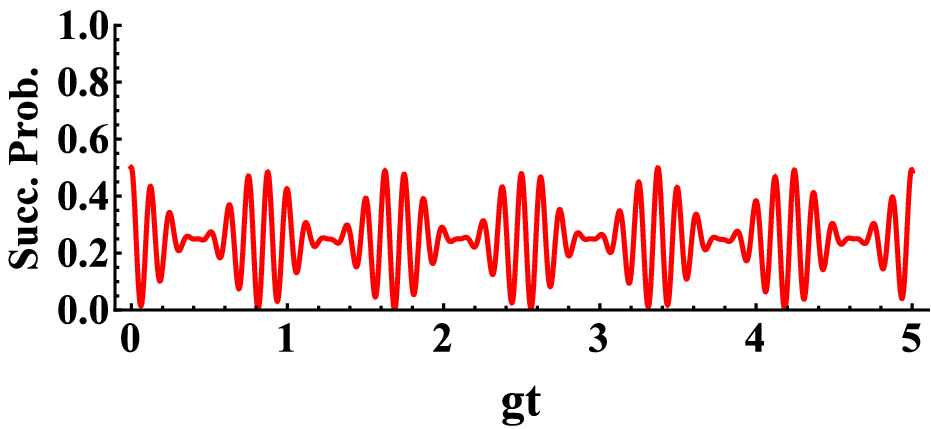}}
               \caption{\label{fig.fig4} {\it Evolution of} (a) concurrence $C^1_{1,4}(t)$ and (b) success probability $P^1_{1,4}(t)$ both related to Eq. (\ref{ent14}) (solid green line), (c) concurrence $C^{1,1}_{1,8}(t)$ and (d) success probability $P^{1,1}_{1,8}(t)$ both related to Eq. (\ref{ent18}) (solid blue line), (e) concurrence $C^{'1,1}_{1,8}(t)$ and (f) success probability $P^{'1,1}_{1,8}(t)$ both related to Eq. (\ref{ent218}) (solid red line) for $\omega_4=\omega_M=0.4g$, $E_P=5g$ with $g_2=g_3=g$, $\nu+\omega_0=\omega_P$, $gt>0$.}
               \end{figure}
 \section{Summary and conclusions}\label{Summary}
 Distribution of entangled state was considered in our study using quantum repeater protocol including trapped atomic ions interacting in OMCs where we have used "interaction"
  as well as "BSM" methods for entanglement swapping needed in the protocol. Eight trapped ions labeled by $(1,2,\cdots,8)$ were considered,
  where the pairs $(i,i+1)$ with $i=1,3,5,7$ were initially prepared in Bell-like states. The entanglement between pairs $(1,4)$ as well as
   $(5,8)$ was produced by performing "interaction" between pairs $(2,3)$ and $(6,7)$ in two separate OMCs and then via
    operating proper measurements on trapped ions $(2,3)$ and $(6,7)$ (these measurements require appropriate projection operators). Finally, the entanglement was swapped between trapped ions $(1,8)$
     after performing the projective measurement with the appropriate Bell-like states on the states of trapped ions (1,4,5,8) in the last step of the introduced protocol. The generated entanglement can be controlled and improved via tuning the pump laser characteristics $(E_P, \omega_P)$, frequency of ionic transition $(\omega_0)$, frequency of the vibrational mode of ions $(\nu)$ and frequency of the optical mode $(\omega_c)$. We found that the success probabilities of entangled states of trapped ions $(1,4)$ (or $(5,8)$)
      are independent of time, detuning and amplitude of pump laser. Also, the intensity of fluctuations of entanglement and success probability
       related to trapped ions $(1,8)$ is decreased by increasing the amount of detuning, \textit{i.e.,} by increasing the mechanical frequency and decreasing $\nu$, $\omega_P$ (note that we have assumed that the frequency of optical mode and the frequency of ionic transition are constant).
       The maxima of entanglement related to target pair $(1,8)$ have been achieved in more times by
        increasing the amount of amplitude of pump laser.
        Also, in this condition, the collapse-revival phenomenon may be appeared for success probability of the entangled state of target trapped ions $(1,8)$.
        Summing up, we may emphasize that, in our protocol we have used the ionic systems and OMCs instead of neutral atoms and
         optical systems. We repeated our calculations in this paper, not shown here, wherein we used optical cavities instead of OMCs and observed that one obtains less entangled states for target trapped ions (1,8). Also, in the presence of OMC the death of entanglement is occurred in less moments of time comparing with using optical cavities. Also, in two of the three processes of measurement, we have performed only appropriate measurements instead of BSM.
         This is another advantage of our proposed model for distributing the entanglement. Since as is well-known the BSM process
          is a hard task to do in this contents.\\
          
  \section{Appendix A: A theoretical scheme to produce the entangled initial state (\ref{initialstates})}\label{Appc}
 In order to generate the Bell-like state (\ref{initialstates}) between two trapped atomic ions, an interaction is performed between them in an optical cavity in the Lamb-Dick regime    as well as the first vibrational sideband \cite{Buzek1997}. The interaction between two trapped ions $(1,2)$ in an optical cavity can be described by the Hamiltonian $\hat{H}=\hat{H}_0+\hat{H}_1$ $(\hbar=1)$, where:
  \begin{eqnarray}\label{hamiltonianopt}
   \hat{H}_0&=&\omega_c\hat{a}^{\dagger}\hat{a}+\sum_{i=1,2}\left(\nu \hat{c}_i^{\dagger}\hat{c}_i+\dfrac{\omega_0}{2}\hat{\sigma}_z^{(i)} \right) ,\\\nonumber
   \hat{H}_1&=&i\sum_{i=1,2} g(\hat{a}\hat{\Sigma}_{+}^{(i)}-\hat{a}^{\dagger}\hat{\Sigma}_{-}^{(i)}).
    \end{eqnarray}
  The involved parameters in (\ref{hamiltonianopt}) have been introduced after Eq. (\ref{hamiltonian}). The effective Hamiltonian using the method mentioned in Appendix B is obtained as below:
 \begin{eqnarray}\label{effopt}
     \hat{H}^\mathrm{eff}_{(1,2)}&=&-\dfrac{g^2}{\delta}\sum_{i=1,2}\hat{\Sigma}_{+}^{(i)}\hat{\Sigma}_{-}^{(i)}-\dfrac{g^2}{\delta}\left( \hat{\Sigma}_{-}^{(1)}\hat{\Sigma}_{+}^{(2)}+\hat{\Sigma}_{-}^{(2)}\hat{\Sigma}_{+}^{(1)}\right),\\\nonumber
     \delta&=&\omega_c-\nu-\omega_0.
\end{eqnarray}
Supposing that the trapped atomic ions (1,2) have been prepared in a separable initial state $\ket{\cal E}_1\otimes\ket{\cal G}_{2}$, the effective Hamiltonian (\ref{effopt}) leads to the following general state
\begin{eqnarray}\label{initialstatesopt}
 \ket{\psi(t)}_{1,2}&=& \cos\left(\dfrac{g^2t}{\delta} \right) \ket{\cal E}_1\ket{\cal G}_{2}+i\sin\left(\dfrac{g^2t}{\delta} \right)\ket{\cal G}_1\ket{\cal E}_{2}.
   \end{eqnarray}
   At a particular moment $t=\dfrac{\pi\delta}{4g^2}$, the time-dependent state (\ref{initialstatesopt}) is converted to the following Bell-like state
\begin{eqnarray}\label{bellopt}
 \ket{\psi(t=\dfrac{\pi\delta}{4g^2})}_{1,2}&=&\dfrac{1}{\sqrt{2}} \left(\ket{\cal E}_1\ket{\cal G}_{2}+i\ket{\cal G}_1\ket{\cal E}_{2}\right),
   \end{eqnarray}
where via applying the phase shift gate and Pauli-$z$ gate \cite{Djordjevic2012}
 \begin{eqnarray}\label{equ24}
S =
\begin{pmatrix}
1 & 0 \\
0 & e^{i\pi/2} \\
\end{pmatrix},\quad
Z =
\begin{pmatrix}
1 & 0 \\
0 & -1 \\
\end{pmatrix}.
\end{eqnarray}
  on the vibrational mode of the trapped ion labeled by 2, the entangled state (\ref{bellopt}) is converted to
 the following Bell-like state,
  \begin{eqnarray}\label{belllike}
  \ket{\psi}_{1,2}&=&\frac{1}{\sqrt{2}}(\ket{\cal E}_1\ket{\cal G}_2+\ket{\cal G}_1\ket{\cal E}_2).
    \end{eqnarray}
  Similarly, the pairs (3,4), (5,6) and (7,8) can be prepared in the Bell-like state (\ref{belllike}).

  \section{Appendix B: Calculating the effective Hamiltonian (\ref{eff})}\label{Appeff}
  The effective Hamiltonian (\ref{eff}) has been obtained using the approach introduced in Ref. \cite{Gamel2010}. At first, the Hamiltonian in the interaction picture is calculated using Hamiltonian (3) and the Baker-Hausdorff lemma as below:
  \begin{eqnarray}\label{intH}
      \hat{H}_\mathrm{int}?(t)&=&e^{i \hat{H}_0 t} \hat{H}_1e^{-i \hat{H}_0 t}\\ \nonumber
      &=&\hat{H}_1+it\left[\hat{H}_0,\hat{H}_1 \right]+\frac{(it)^2}{2!} \left[\hat{H}_0,\left[\hat{H}_0,\hat{H}_1 \right]\right]+\cdots
       \end{eqnarray}
  For instance, the two terms of (\ref{intH}) are respectively obtained straightforwardly as below:
      \begin{eqnarray}\label{h1}
      \left[\hat{H}_0,\hat{H}_1 \right]&=&-G \hat{a}^\dagger\hat{a} \omega_M (\hat{b}^\dagger-\hat{b})-i(\omega_c-\nu-\omega_0)\sum_{i=2,3}g_i(\hat{a}\hat{\Sigma}^{(i)}_++\hat{a}^\dagger\hat{\Sigma}^{(i)}_-)\\\nonumber
      &+& iE_P\omega_c(\hat{a}^\dagger e^{-i \omega_P t}+\hat{a} e^{i \omega_P t}),
         \end{eqnarray}
  and
      \begin{eqnarray}\label{h21}
      \left[\hat{H}_0,[\hat{H}_0,\hat{H}_1 ] \right]&=&-G \hat{a}^\dagger\hat{a} \omega^2_M (\hat{b}^\dagger+\hat{b})+i(\omega_c-\nu-\omega_0)^2\sum_{i=2,3}g_i(\hat{a}\hat{\Sigma}^{(i)}_+-\hat{a}^\dagger\hat{\Sigma}^{(i)}_-)\\\nonumber
          &+& iE_P\omega^2_c(\hat{a}^\dagger e^{-i \omega_P t}-\hat{a} e^{i \omega_P t}).
      \end{eqnarray}
      By substituting Eqs. (\ref{h1}), (\ref{h21}) into the formula (\ref{intH}), one can obtain the following interaction Hamiltonian in the interaction picture
    \begin{eqnarray}\label{hbl}
       \hat{H}^\mathrm{int}_{(2,3)}&=&-G\hat{a}^{\dagger}\hat{a}(\hat{b}e^{-i \omega_1 t}+\hat{b}^{\dagger}e^{i \omega_1 t})-i g_2\left( \hat{a}^{\dagger}\hat{\Sigma}_{-}^{(2)} e^{i\omega_2t}-\hat{a}\hat{\Sigma}_{+}^{(2)}e^{-i\omega_2t}\right) \\ \nonumber
      &-&i g_3\left( \hat{a}^{\dagger}\hat{\Sigma}_{-}^{(3)} e^{i\omega_3t}-\hat{a}\hat{\Sigma}_{+}^{(3)}e^{-i\omega_3t}\right) -i E_P \left(\hat{a} e^{-i \omega_4 t}-\hat{a}^{\dagger} e^{i \omega_4 t}\right),
           \end{eqnarray}
           where $\omega_1=\omega_M$, $\omega_2=\omega_3=\omega_c-\nu-\omega_0$, $\omega_4=\omega_c-\omega_P$. Now, as proved in Ref. \cite{Gamel2010}, when the Hamiltonian in the interaction picture is as below:
            ?\begin{eqnarray}\label{int}
            \hat{H}_{\mathrm{?int}}(t)=\sum_{n=1}^{N}\hat{h}_n \exp({-i\omega_{n}t})+\hat{h}_n^{\dagger} \exp({i\omega_{n}t})?,
            \end{eqnarray}
            then, the effective Hamiltonian can be achieved as,
           ?\begin{eqnarray}\label{eff2}
           \hat{H}_{\mathrm{eff}}(t)=\sum_{m,n=1}^{N}{1\over\hbar{\omega}_{{mn}}}[\hat{h}_m^{\dagger},\hat{h}_n] \exp\left({i\left[\omega_{m}-\omega_{n}\right]}t\right),
           ?\end{eqnarray}
   where N is the total number of different harmonic terms
   making up the interaction Hamiltonian, $\omega_{n}>0$ is
   frequency and ${\omega}_{{mn}}$ is the harmonic average of $\omega_{m}$ and $\omega_{n}$ in the
   sense that,
   ?\begin{eqnarray}\label{A13}?
   ?{1\over{\omega}_{{mn}}}={1\over2}\left({1\over\omega_m}+{1\over\omega_n}\right).
   ?\end{eqnarray}???
   Accordingly, in our case, the effective Hamiltonian using Eqs. (\ref{hbl}), (\ref{int}), (\ref{eff2}) and (\ref{A13}) can be obtained as below:
 \begingroup\makeatletter\def\f@size{6}\check@mathfonts
  \begin{eqnarray}\label{eff22}
  \tiny
       \hat{H}^\mathrm{eff}_{(2,3)}&=&-\dfrac{g^2_2}{\omega_{22}}\hat{\Sigma}_{+}^{(2)}\hat{\Sigma}_{-}^{(2)}-\dfrac{g^2_3}{\omega_{33}}\hat{\Sigma}_{+}^{(3)}\hat{\Sigma}_{-}^{(3)}-\dfrac{g_2 g_3}{\omega_{23}}\left( \hat{\Sigma}_{-}^{(2)}\hat{\Sigma}_{+}^{(3)}e^{i(\omega_2-\omega_3)t}+\hat{\Sigma}_{-}^{(3)}\hat{\Sigma}_{+}^{(2)}e^{-i(\omega_2-\omega_3)t}\right)\\\nonumber
       &+&\dfrac{g_2 E_P}{\omega_{24}}\left( \hat{\Sigma}_{+}^{(2)}e^{i(\omega_4-\omega_2)t}+\hat{\Sigma}_{-}^{(2)}e^{-i(\omega_4-\omega_2)t}\right)+\dfrac{g_3 E_P}{\omega_{34}}\left( \hat{\Sigma}_{+}^{(3)}e^{i(\omega_4-\omega_3)t}+\hat{\Sigma}_{-}^{(3)}e^{-i(\omega_4-\omega_3)t}\right)\\\nonumber
       &-&\dfrac{G^2}{\omega_{11}}(\hat{a}^\dagger \hat{a})^2+iG\left(\hat{a}\hat{b}^\dagger \sum_{i=2,3} \dfrac{g_i}{\omega_{1i}} \hat{\Sigma}_{+}^{(i)} e^{i(\omega_1-\omega_i)t}-H.C.\right)+i \dfrac{E_P G} {\omega_{14}} \left( \hat{a}^\dagger\hat{b}e^{i(\omega_4-\omega_1)t}-\hat{a}\hat{b}^\dagger e^{i(\omega_1-\omega_4)t}\right)\\\nonumber
       &-&\hat{a}^\dagger\hat{a}\sum_{i=2,3}\dfrac{g^2_i}{\omega_{ii}}\hat{\Sigma}_{z}^{(i)} ,
           \end{eqnarray}
   where $\hat{\Sigma}_{z}^{(i)}=\left[\hat{\Sigma}_{+}^{(i)},\hat{\Sigma}_{-}^{(i)} \right] $ and $\dfrac{1}{\omega_{ij}}=\dfrac{1}{2}(\dfrac{1}{\omega_i}+\dfrac{1}{\omega_j})$. By consideration of initial vacuum states for optical and mechanical modes, the effective Hamiltonian (\ref{eff22}) reduces to Eq. (\ref{eff}). Notice that the frequencies $\omega_n$ in Eq. (\ref{int})  should be close to each other \cite{Gamel2010}.
         \section{Appendix C: Calculating the coefficients of entangled state (\ref{state1-4})}\label{App}
          In this section, the differential equations related to entangled state (\ref{state1-4}) using the effective Hamiltonian (\ref{eff}) and time-dependent Schr\"{o}dinger equation $i\frac{\partial }{\partial t}\ket{\psi(t)}=?\hat{H}^\mathrm{eff}_{(2,3)}\ket{\psi(t)}$ are achieved as below
          \begin{eqnarray}\label{lasteq}
           i \dot{X}(t)=S(t) X(t),
          \end{eqnarray}
          where $X(t)$ and $S(t)$ are respectively defined as follow,
         \begin{eqnarray}
              \tiny
                X(t)&=& \left( \begin{array}{cccc}
                    x_{1}(t)\\
                    x_{2}(t)\\
                    x_{3}(t)\\
                    x_{4}(t)\\
                   \end{array} \right),
                 \label{eq:unitarymatrix1}
                 \end{eqnarray}
                  \begin{equation}
                                         S(t)= \left( \begin{array}{cccc}
                      -\dfrac{g^2_3}{\omega_{33}}  & -\dfrac{g_2g_3}{\omega_{23}}e^{i(\omega_2-\omega_3)t} & \dfrac{g_2E_P}{\omega_{24}}e^{-i(\omega_4-\omega_2)t} & \dfrac{g_3E_P}{\omega_{34}}e^{i(\omega_4-\omega_3)t}\\
                      -\dfrac{g_2g_3}{\omega_{23}}e^{-i(\omega_2-\omega_3)t}  & -\dfrac{g^2_2}{\omega_{22}} & \dfrac{g_3E_P}{\omega_{34}}e^{-i(\omega_4-\omega_3)t} & \dfrac{g_2E_P}{\omega_{24}}e^{i(\omega_4-\omega_2)t}\\
                       \dfrac{g_2E_P}{\omega_{24}}e^{i(\omega_4-\omega_2)t}  & \dfrac{g_3E_P}{\omega_{34}}e^{i(\omega_4-\omega_3)t} & -\left( \dfrac{g^2_2}{\omega_{22}}+\dfrac{g^2_3}{\omega_{33}}\right) & 0\\
                       \dfrac{g_3E_P}{\omega_{34}}e^{-i(\omega_4-\omega_3)t}  & \dfrac{g_2E_P}{\omega_{24}}e^{-i(\omega_4-\omega_2)t} & 0 & 0 \\
                        \end{array} \right),
                     \label{eq:unitarymatrix0}
                     \end{equation}
       where $x_i(t)\equiv\alpha_i(t),\beta_i(t),\gamma_i(t),\eta_i(t)$ with $i=1,2,3,4$, each of $\alpha_i(t),\beta_i(t), \cdots$, contains four differential equations, \textit{i.e.,} summing up, one obtains sixteen differential equations.
       The Eq. (\ref{lasteq}) has been solved using Laplace transform techniques with the help of Mathematica software by considering $\omega_4=\omega_M$ and $\nu+\omega_0=\omega_P$, \textit{i.e.,} $\omega_1=\omega_2=\omega_3=\omega_4=\omega_M$. In fact, these assumptions are required to calculate the effective Hamiltonian (\ref{eff}), the conditions that Eq. (\ref{lasteq}) may be easily derived.

\end{document}